\documentclass[letterpaper, 10 pt, conference]{ieeeconf}  % Comment this line out if you need a4paper

\IEEEoverridecommandlockouts                              % This command is only needed if 
\usepackage{graphicx}
\usepackage{cite}
\usepackage{url}
\usepackage{subcaption}
\usepackage{float}
\title{\LARGE \bf
A Smart Security System with Face Recognition 
}

\author{Trung Nguyen, Barth Lakshmanan and Weihua Sheng$^{1}$ %\IEEEmembership{Senior Member, IEEE}% <-this % stops a space
\thanks{$^{1}$Trung Nguyen, Barth Lakshmanan and Weihua Sheng are with the Laboratory for Advanced Sensing, Computation and Control (ASCC Lab), School of Electrical and Computer Engineering, Oklahoma State University, Stillwater, OK, 74078, 
	USA.
        {\tt\small trungdn@okstate.edu, bharal@okstate.edu, weihua.sheng@okstate.edu }}%
}

\begin{document}

\maketitle
\thispagestyle{empty}
\pagestyle{empty}

%%%%%%%%%%%%%%%%%%%%%%%%%%%%%%%%%%%%%%%%%%%%%%%%%%%%%%%%%%%%%%%%%%%%%%%%%%%%%%%%
\begin{abstract}

Web-based technology has improved drastically in the past decade. As a result, security technology has become a major help to protect our daily life. In this paper, we propose a robust security based on face recognition system (SoF). In particular, we develop this system to giving access into a home for authenticated users. The classifier is trained by using a new adaptive learning method. The training data are initially collected from social networks. The accuracy of the classifier is incrementally improved as the user starts using the system. A novel method has been introduced to improve the classifier model by human interaction and social media. By using a deep learning framework - TensorFlow, it will be easy to reuse the framework to adopt with many devices and applications. 

\end{abstract}

%%%%%%%%%%%%%%%%%%%%%%%%%%%%%%%%%%%%%%%%%%%%%%%%%%%%%%%%%%%%%%%%%%%%%%%%%%%%%%%%
\section{INTRODUCTION}
\subsection{Motivation}

Modernization is leading to a remarkable increase number of crimes, especially robbery. In the report, the law enforcement agencies throughout the US showed an overall increase of 1.7 percent in the number of violent crimes, which are brought to their attention for the first 6 months of 2015; and, robbery has been increased by 1 percent from 311,936 cases in 2014 \cite{crime}. Therefore, Security systems have a crucial role to safeguard people. It is necessary to have a robust system which can distinguish between people and respond differently based on their privileges.

A number of methods are available for detecting and recognizing faces with various levels of complexities. Face recognition facilitates automation and security. It has already been used in many applications including ID issuance, law enforcement, border control, and many other commercial products. The state-of-art recognizers using convolutional neural networks (CNN) outperform the human's recognition rate; however, these systems are not automatically improving. Another issue with these systems is that it requires adequate data to be trained before it is actually being deployed. It is essential that the system is robust to recognize people and that the training should be accomplished without much difficulty. 

And with Google Brain's second-generation system, TensorFlow \cite{abadi2016tensorflow} is a deep learning framework released by Google. TensorFlow is flexible, portable, feasible, and completely open source. It also extensively interacts with different hardware such as smartphones, and embedded computers \cite{tran2018real}. With the advancement in mobile, cloud and embedded computing, we develop the Security based on Face Recognition System (SoF).
\subsection{Related Work}
Home security has been an essential feature in smart home and received a growing interest in recent years \cite{do2018rish} \cite{do2015developing}. Various home security systems have been used in the market for many pre-potent companies such as ADT \cite{ADT}, Vivint \cite{vivint}, and Protect America \cite{proamerica}. However, none of them have the face-recognition feature in their systems because of moderate confidence and exhausted computational requirements. Along with the rapid development of smart devices, Netatmo \cite{denmead2013netatmo} presented a device using the deep neural network to realize the face, but their system is still far from expectation. This smart camera does not keep up with the competition as a webcam or a security cam. It is slow at everything; the live feed is lagged; the notifications are delayed, and it takes a while to learn faces.

In research problems, there are several security systems using face recognition technology. Facial recognition system for security access and identification presented by Jeffrey S. Coffin \cite{coffin1999facial} uses custom VLSI Hardware and Eigenspaces method, and security systems presented by Shankar Kartik also uses Eigenfaces method for face identification  \cite{kartik2013security} which gives unfavorable results with moderate accuracy. We use a deep learning algorithm for face recognition problems which is closing the gap to human-level performance in face verification. 

The robustness of face recognition systems depend on the changes in conditions of light or expression or even in the partial blocked of the face can be considered. Several papers have proposed various techniques for face recognition under those conditions. Eigenfaces \cite{turk1991eigenfaces} are variant extracted feature to above factors. Facenet using deep convolutional neural network \cite{schroff2015facenet} with the architecture of Inception model \cite{szegedy2015going} from Google and uses a novel online triplet mining method to train instead of an intermediate bottleneck layer. On the widely used Labeled Faces in the Wild (LFW) dataset \cite{labeled}, Facenet system achieved a new record accuracy of 99.63\%. However, unfortunately, not only the size of the database increases but also its computational cost increases and recognition accuracy declines accordingly. That is why incremental learning is a learning algorithm which approaches to handle large-scale training data to efficiency and accuracy. A brief definition of incremental learning is that learning is a gradual process with new data. The idea is the existed classifiers that are identified with the new classes to be learned \cite{kuzborskij2013n}. Its key idea is to begin learning on low-resolution images and then gradually increase to high-resolution image \cite{xiao2014error}. We use 96x96 pixels cropped as the input data which is also mentioned in Facenet \cite{schroff2015facenet} for training data size. 

Face detection algorithms have been addressed in many papers such as Haar Cascades \cite{wilson2006facial}, Kalman Filter \cite{qian1998robust} and applied in various fields \cite{do2014human} \cite{sheng2013integrated} \cite{tran2016driver}. Besides, OpenCV is a robust open source which supports many methods to detect faces. However, in this paper, we use the Dlib C++ library which supports machine learning algorithms and uses histogram-of-oriented-gradient algorithm \cite{king2009dlib}. Face detection is not only necessary for the camera node detecting the face but also helpful in pre-processing the input data. 
In this paper, we also describe a novel method to collect the data from social media by using Facebook API and ask from human interaction to label the unidentified peoples, that directly incremental learn for the neural network model with new data. The interface is also designed with easy uses in many types of devices.

\section{AN OVERVIEW OF OUR SYSTEM}

This section describes SoF system experimental architecture inside a smart home. The SoF system consists of a camera node, a cloud server, a smartphone and smart devices for interacting with users. The SoF system is shown in Figure 1.

\begin{figure}[!h]
	\centering
	\includegraphics[width=3.3in]{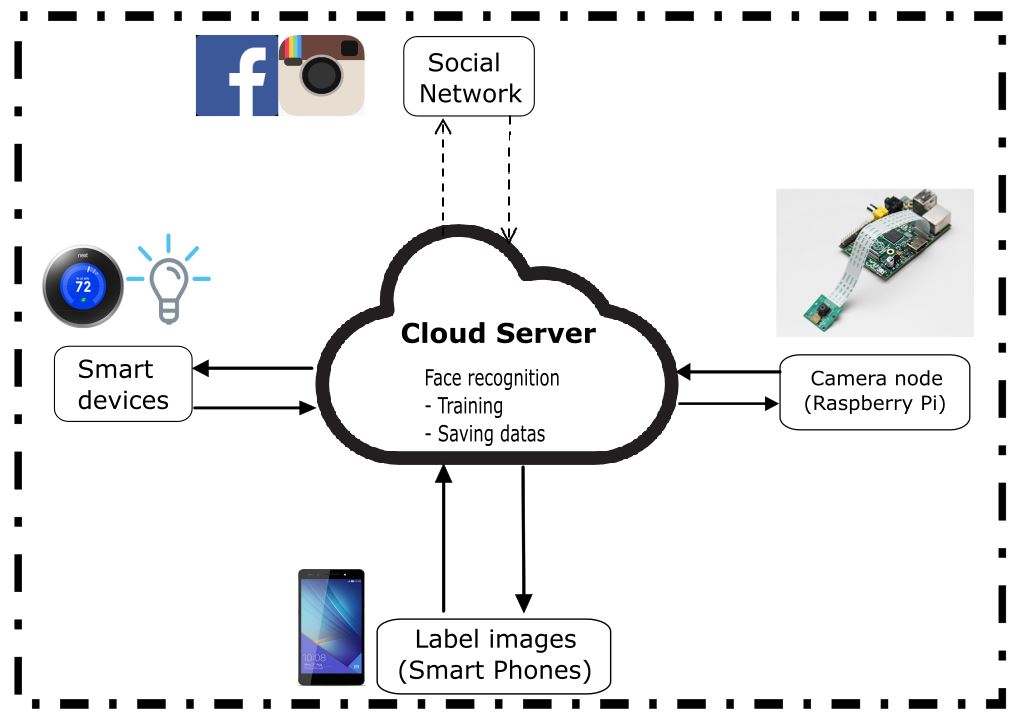}
	\caption{Home Security System Architecture.}
	\label{webpg}
\end{figure}

\subsection{Camera Node}

The camera node uses a Raspberry Pi, a tiny and affordable computer, which is typically placed near the entrance where the access has to be granted. Whenever a person needs access to the house, the camera node will capture a photo, and process it further. The camera nodes are positioned such that it has a wide range of vicinity over the subject and it can detect the face approaching the camera from distance. 

The camera node first will detect the human face, and then it directly runs the all image processing off-line, locally in a Raspberry Pi by using Dlib library and TensorFlow installed inside Raspberry Pi. 

Raspberry Pi is a tiny embedded computer with limited power, the training neural network requires expensive computation, so the training task will be done in the cloud node.

In addition, the camera node can adaptively be a smartphone or a sensor camera node or an assistant robot because TensorFlow is able to run on many operation systems. 

\subsection{Cloud Server}

Cloud computing has found a drastic advancement recently. Cloud computing is a type of computing that relies on sharing computing resources rather than having local servers or personal devices to handle applications. Cloud computing provides a simple way to access servers, storage, databases and a broad set of application services for research over the internet. A face recognition application using CNN requires a lot of computational power machine or computer which will need general purpose GPUs. Cloud computing offers a reliable solution at low cost for such kind of applications.
Following on the architecture, the cloud server node will receive data from the camera nodes and save then train the data after collection period. It also interacts with the owner/administrator through a smart device. The server has a database with a record of all users. The server could communicate with the sensor nodes and the smart device using web-socket enabling real-time data processing.
The cloud server also has a web-based server collecting data from Facebook and saving all data to storage. Based on the (CoSHE) \cite{pham2018delivering}, \cite{pham2016cloud} cloud infrastructure and cloud testbed for research and education \cite{tran2015cloud}. We built the cloud services for SoF system. The cloud server architecture is shown in Figure 2.

\begin{figure}[!h]
	\centering
	\includegraphics[width=3.3in]{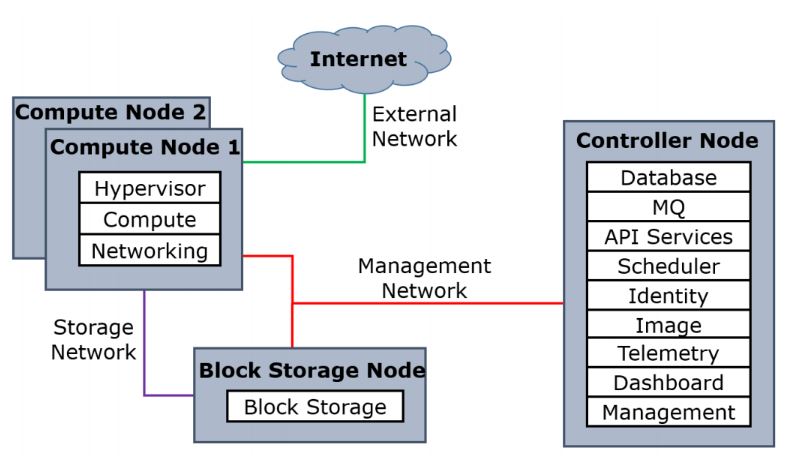}
	\caption{Stack Cloud Architecture.}
	\label{webpg}
\end{figure}

\subsection{Smart Devices}
The cloud server could communicate with smartphones and other smart devices such as smart thermostats, lights, and sensors. These smart devices are controlled by the cloud server.  The smartphone allows the owner to control the smart devices and also to change the permission level for different users. Based on the granted access level, different users will able to control different smart devices. We demonstrate the capabilities of the system using a miniature smart home as shown in Figure 3 \cite{nguyen2017miniature}. Whenever a new person is detected, the cloud server sends an alert to the smartphone. The owner can then label the person name or take necessary actions in case of any security breach. Also, smart devices are dramatically growing in recent years, it is much more convenient to control smart devices through IoT (Internet of Things).
\begin{figure}[!h]
	\centering
	\includegraphics[width=3.3in]{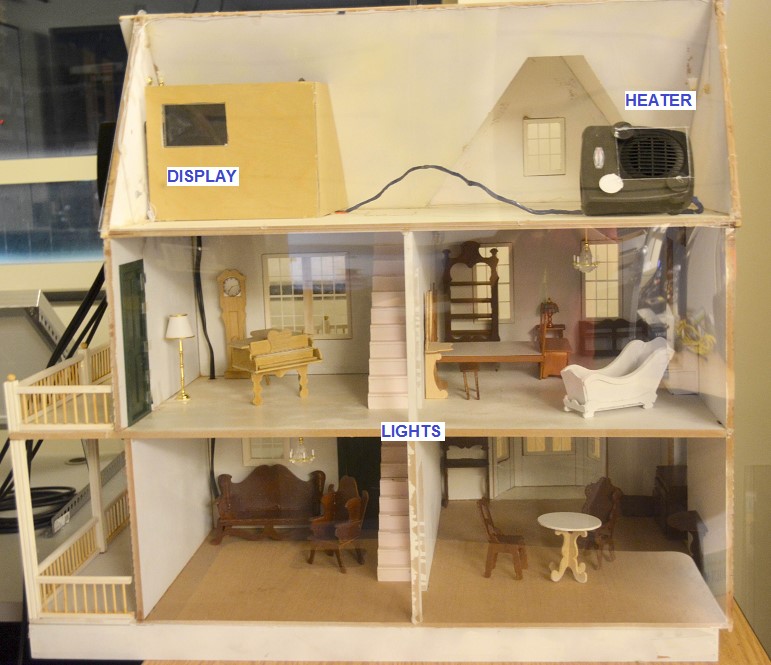}
	\caption{Miniature Smart Home.}
	\label{webpg}
\end{figure}

\subsection{Social Network}

This is a new approach to collect data. Social networks are the largest free, diversified, adaptive data online. By using the advantage of Graph API from Facebook, we could easily detect a face with a tag name. Then, we could download all picture with user's faces to the cloud node. And most importantly, the facebook developer app is simple and convenient to share between users. They only need to log-in with their account and accept the app to collect pictures. The social network node has three collecting interfaces. First is a public application from Facebook developer website, second is from a web-based hosted on the cloud server, the last one is an application on Android devices. These are easy to collect face images with labeled faces of users who are given access. We also mentioned that data will be used for research purpose and protected sensitive data for users. 

\section{TRAINING AND INCREMENTAL LEARNING}

\subsection{Introduction} 

The human brain makes vision seem very easy. It does not take any difficulty to tell apart a cheetah and a tiger, read a sign or recognizes a human face. But these are really difficult problems to solve with a computer. They only seem easy because human brains are fabulously good at understanding images. In recent years, machine learning has made marvelous progress in solving these difficult problems. In particular, the model called a deep convolutional neural network can achieve reasonable performance on difficult visual recognition tasks which are matching or exceeding human performance in some domains.
Researchers have demonstrated reliable methods in computer vision by validating their work in ImageNet \cite{deng2009imagenet} - an academic benchmark for computer vision. Subsequent models improve each time to achieve a new state-of-the-art result: QuocNet \cite{le2013building}, AlexNet \cite{krizhevsky2012imagenet}, Inception (GoogLeNet), BN-Inception-v2 and Inception-v3 \cite{szegedy2015rethinking}. Inception-v3 is the latest trained model for the ImageNet Large Visual Recognition Challenge from Google.
We implemented the face recognition module based on the method presented in Facenet \cite{schroff2015facenet} and the training inception-v3 model in TensorFlow \cite{abadi2016tensorflow}. Instead of using Inception (GoogLeNet) model architecture, we use Inception-v3 architecture to train a new model with improved accuracy. 

\subsection{Architecture}

Based on the architecture and published model from the Openface \cite{amos2016openface} and Inception-v3 model \\cite{abadi2016tensorflow}. We trained a new model with a new database set.
The face recognition architecture is shown in Figure 4. The input data is aligned by using a face detection method and then goes through the deep convolutional neural network to extract an embedding feature. We can use the feature for similarity detection and classification. 

\begin{figure}[!h]
	\centering
	\includegraphics[width=3.5in]{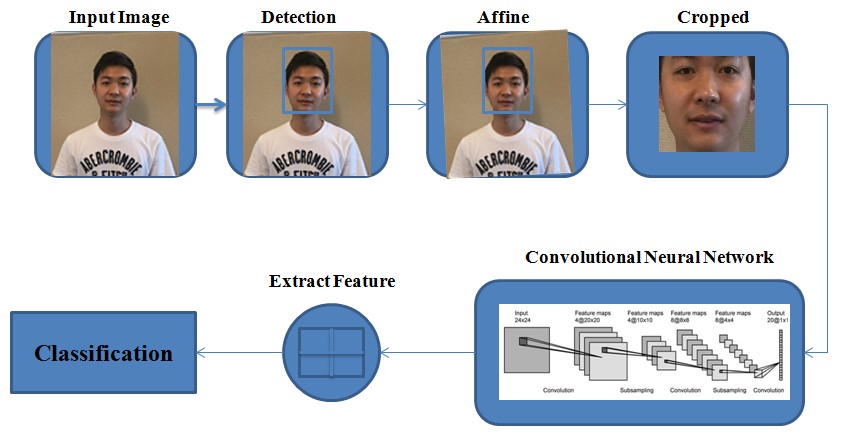}
	\caption{Face Recognition Architecture.}
	\label{webpg}
\end{figure}

\subsection{Detection and Affine Transformation}

When processing an image, face detection (Dlib library) \cite{king2009dlib} first find a square around faces. Each face is then passed separately into the neural network, which expects a fixed-sized input, currently 96x96 pixels as mentioned in Facenet \cite{schroff2015facenet}, which is the best size giving the highest accuracy and low training time. We reshape the face in the square to 96x96 pixels. A potential issue with this is that faces could be looking in different directions and we have to rotate the images. We use align faces method described in OpenFace \cite{amos2016openface}  by first finding the locations of the eyes and nose with Dlib's landmark detector, and if the face is undetected or unaligned which will be eliminated before going to the neural network. Finally, an affine transformation is performed on the cropped image to make the eyes and nose appear at about the same place as in Figure 5.

\begin{figure}[!h]
	\centering
	\includegraphics[width=3.3in]{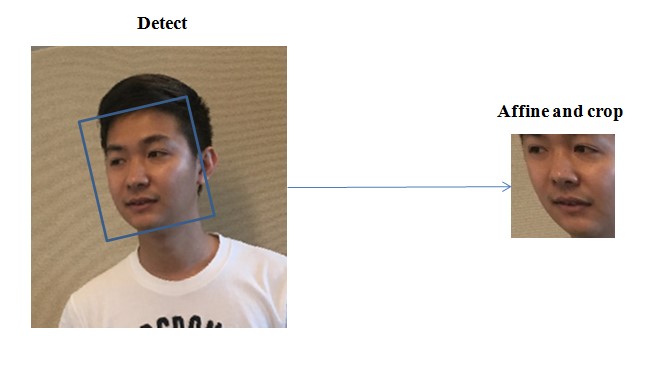}
	\caption{Cropped and Affined data.}
	\label{webpg}
\end{figure}

\subsection{Initial Training Using Data from Social Network}

Our model trained 2622 celebrities from the VGG-Face dataset \cite{Parkhi15}, 402 people from Facebook and 108 students using the security system. 
The process of collecting data of users and training the data is somewhat cumbersome. In order to make the training process easier, we obtain the data from different social network accounts. Images of new users are obtained from the social media once the owner requests access for a particular user using the smartphone and new users after using the security system. 

\subsection{Incremental Learning}

Often the training data obtained from the social network are insufficient for the deep learning model to perform accurately. Once the user is trained with a minimal dataset from the social media, the representation of the user is further improved by fine-tuning the model as the user starts using the system. Sometimes the face recognition system fails to classify the person properly and will have a very low accuracy, in such case; the system asks help from the owner. The owner is sent a request to label the person through his smart phone. After the owner has labeled, the system will automatically update with the new data and send back to the camera node to give the access. The interface is also built in a website and an android app which are friendly to label and collect data. We use the Triplet loss method mentioned in Facenet \cite{schroff2015facenet} for incremental learning.  

\section{SYSTEM WORKING AND COLLECTING DATA}

\subsection{System working}

The system has two processing nodes. The first one is from the camera node as in Figure 6. By using a Raspberry Pi with Pi camera, Raspberry Pi will detect and realize the human face with the current model and data stored in memory. Giving the access or send data to the server is based on if the system is able to detect and recognize the face with set-up confidence. If the confidence is low or unable to recognize the face, the Raspberry Pi will take a series of user's photo with different angles and expression to store in the cloud for training purpose. After the training task done in the cloud, the updated version of the new model will download to the Raspberry Pi.

\begin{figure}[!h]
	\centering
	\includegraphics[width=3.3in]{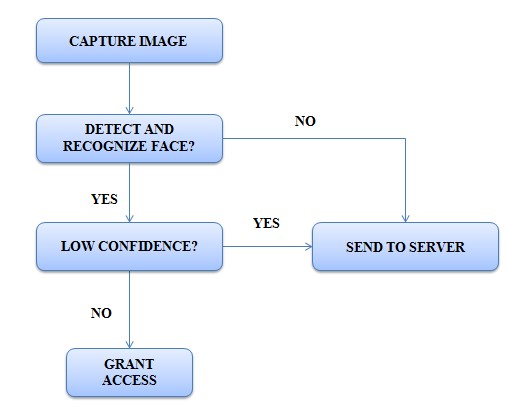}
	\caption{Camera node processes.}
	\label{webpg}
\end{figure}

The second process is shown in Figure 7 in the cloud node. Cloud node aimed to store the face data and send alerts to the owner asking for labeling the unidentified person. In addition, the Facebook web-based application is built in the cloud server, which collects the data from social media. The cloud server will do the much exhausted computational training tasks. By using the distributed TensorFlow, we trained the model in multiple computing nodes to speed up the training time and also using the incremental learning technique in Facnet \cite{schroff2015facenet} to retrain the model with new data collected from social networks and the security systems after specified time uses.  

\begin{figure}[!h]
	\centering
	\includegraphics[width=3.3in]{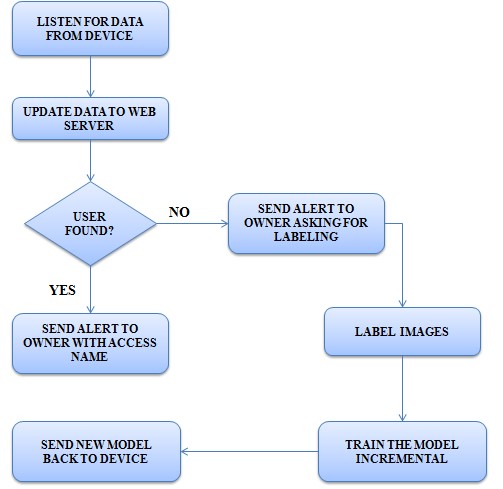}
	\caption{Cloud node processes.}
	\label{webpg}
\end{figure}

\subsection{Collecting data from Facebook}

The biggest problem of deep neural network is data. As we mentioned in section III, the VGG dataset is only around 2.6 million images with around 2.6 thousand identities. To compare with Google datasets mentioned in Facenet paper [\cite{schroff2015facenet}, they use hundreds of millions of images from Google and Youtube. On the purpose of researching or business, you have to pay for a robust face dataset or manually collecting the data will take a while, and the data are also insufficient. However, as social media becomes more popular around the world, we proposed a novel method to automatically collect the data from social media. In this paper, we only mentioned Facebook since it is the largest social network today, but actually, we can use this method in other social media networks. 
By using the graph API for developers, we can extract the tag face from the users by giving the login access. We also built and published an application on Facebook which is very convenient for all users around the world who can log-in and share their face images. The working flow of Facebook Graph API is shown in Figure 8.

\begin{figure}[!h]
	\centering
	\includegraphics[width=3.5in]{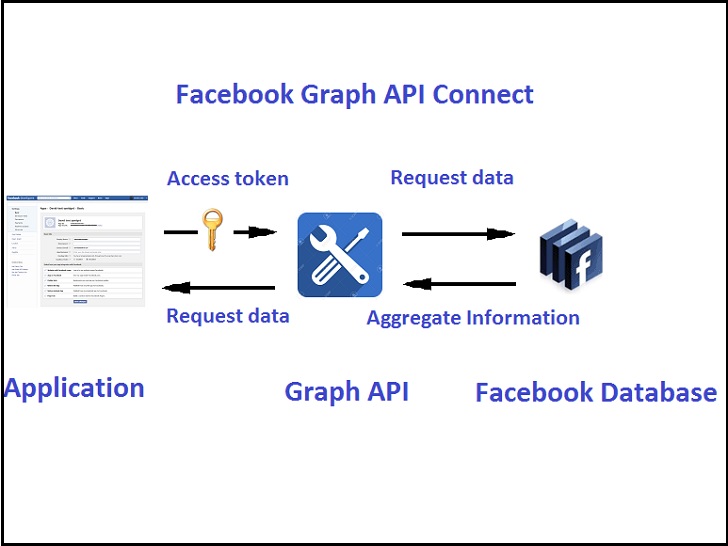}
	\caption{Facebook graph API.}
	\label{webpg}
\end{figure}

\section{EXPERIMENTAL RESULTS}
\subsection{Face recognition performance}

First, we tested SoF model which trained by VGG dataset \cite{Parkhi15} on Labeled Faces in the Wild (LFW) datasets \cite{labeled} and the classification accuracy is 0.9318 $\pm$ 0.0140. The ROC of SoF model is shown in Figure 9 compared with Human and Eigenfaces experiments. Unfortunately, the model is unable to reach the accuracy mentioned in the Facenet paper \cite{schroff2015facenet} since using fewer input data to compare with billion photos from Google. and also we use a different method to pre-process the input data. However, the accuracy is obviously impressed to compare with Eigenfaces algorithm used in Jeffrey's \cite{coffin1999facial} and Shankar's \cite{kartik2013security} security system. The state-of-art Inception-v3 model gives a marvelous result which closes the human gap. 

\begin{figure}[!h]
	\centering
	\includegraphics[width=3.5in]{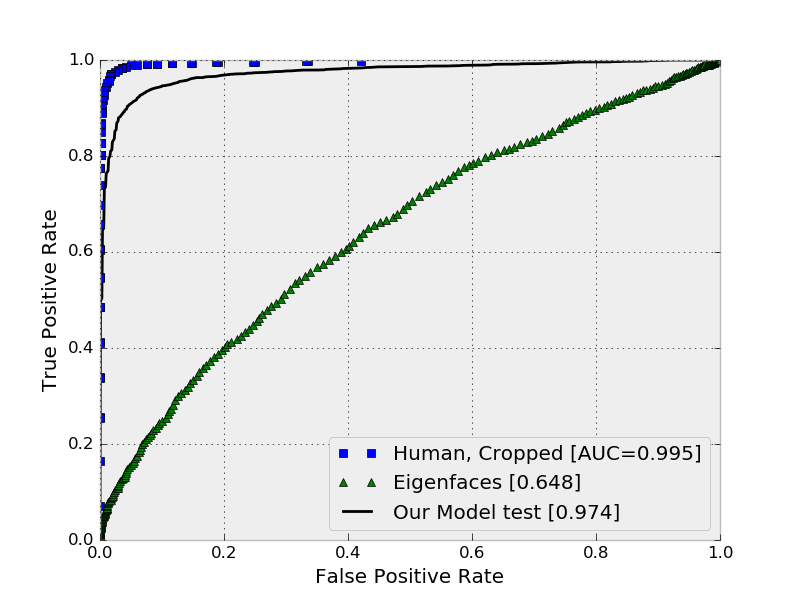}
	\caption{SoF model test ROC with LFW.}
	\label{webpg}
\end{figure}

We also tested face recognition in a real environment by testing 20 people, shown in Figure 10. The highest accuracy is 92.2\%. These people are presented as new guests coming to the house and asking for a label from the owner.  The datasets we used to train the neural network model is American, but 60\% of test data is Asian and Latino. The result was lower accuracy to compare with LFW dataset and it sometimes failed to distinguish people. That makes collecting data from Social Network advantages because it is able to collect various, diverse people from around the world. The SoF systems also confused between two people with similar faces, but more face images with different angles and expressions will solve the problem. The light condition is also important, the background should not be too illuminated. 

\begin{figure}[!h]
	\centering
	\includegraphics[width=3.5in]{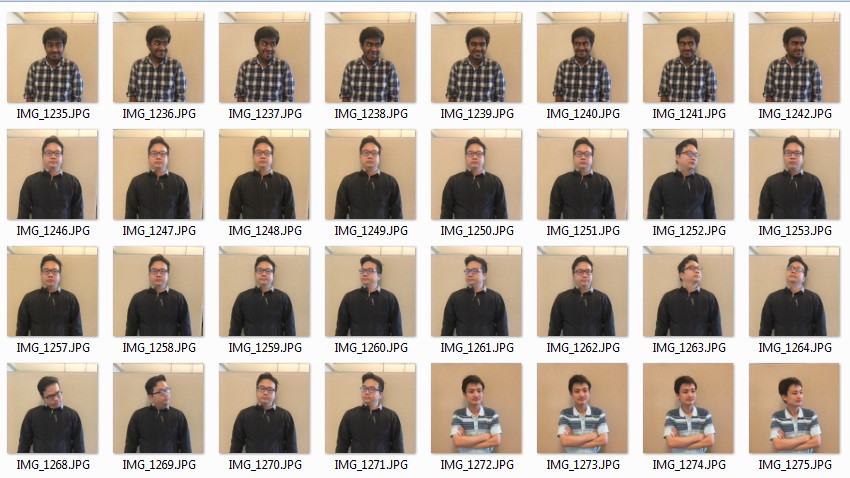}
	\caption{Lab members dataset.}
	\label{webpg}
\end{figure}

The bottleneck values is shown Figure 11 before training to distinguish different faces.  

\begin{figure}[!h]
	\centering
	\includegraphics[width=3.3in]{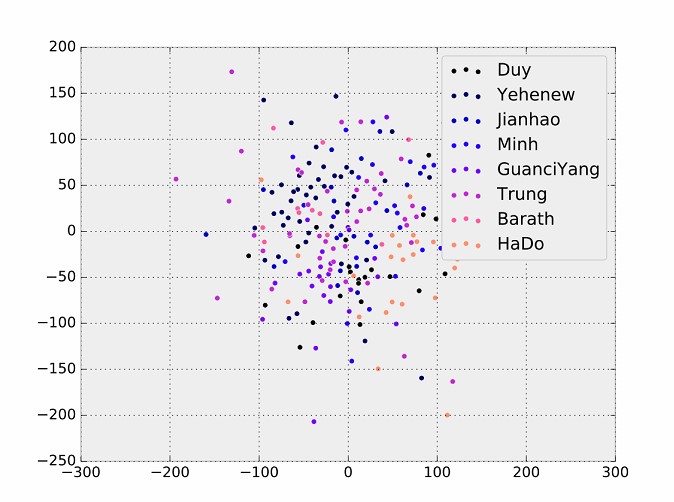}
	\caption{Bottleneck values.}
	\label{webpg}
\end{figure}

After collecting the data from 108 students and 402 users of Facebook, we trained a new model with updated data using incremental learning. The result is shown in Figure 12 with improving AUC from 0.974 to 0.985. Also, the correctly recognize faces of Asian face is increased because we updated the data with diverse images of people from different regions. We would not expect the accuracy will dramatically increase because the data we collected is insufficient and also at the limit of the algorithm. However, by using that incremental learning method, we will reach the accuracy mentioned in the Facenet paper \cite{schroff2015facenet}. We have already improved the accuracy by using the Inception-v3 \cite{szegedy2015rethinking} model. If we focus more on pre-processing the input data by aligning the data and using the TF-Slim libraries with the lightweight package for defining, training and evaluating models, we can even improve the performance more.  

\begin{figure}[!h]
	\centering
	\includegraphics[width=3.3in]{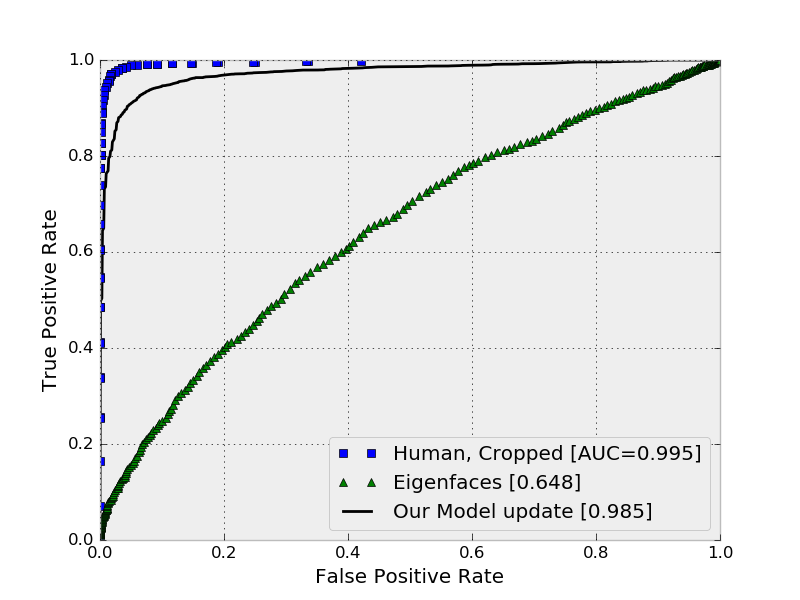}
	\caption{SoF model update ROC.}
	\label{webpg}
\end{figure}

\subsection{Security system setup}

The entire system was developed and tested in a miniature smart home mimicking the actual smart home \cite{nguyen2017miniature}. Raspberry Pi is plugged in the front door. It is always running faces detection, then faces recognition locally. The miniature smart home with raspberry pi is shown in Figure 13.

\begin{figure}[!h]
	\centering
	\includegraphics[width=3.3in]{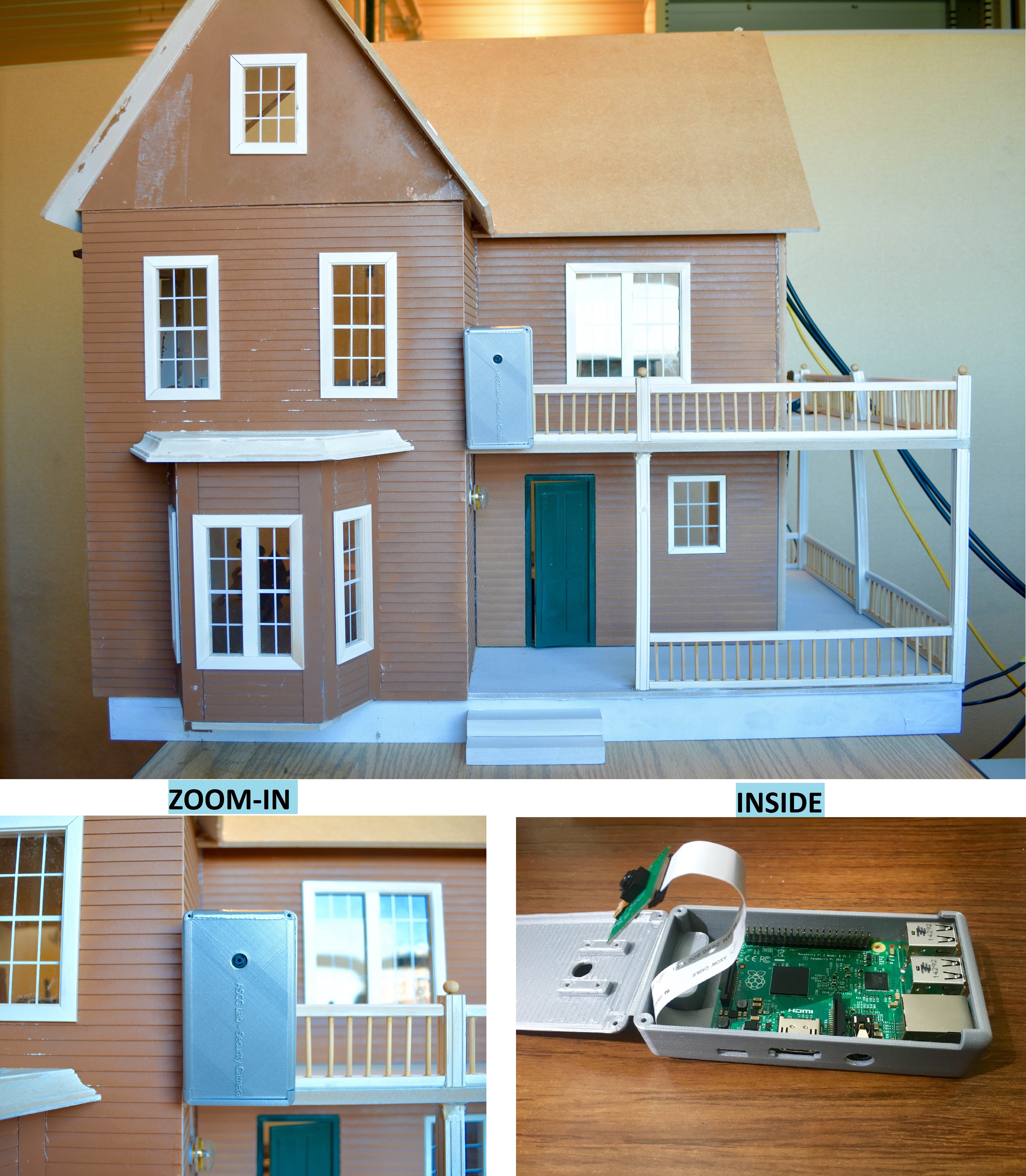}
	\caption{Camera Node (Raspberry Pi).}
	\label{webpg}
\end{figure}

The cloud server is a stack server, which is described in section II. The server includes 3 nodes: 1 controller node on the top and 2 compute nodes in the bottom. We actually run several systems in the cloud with different projects, which make it is easy to share the data and information. Our cloud server includes Ubuntu, Cirros, and CentOs operation environments and instances hosted by the QEMU hypervisor \cite{qemu} on the compute nodes. The cloud server is able maximum to 12 VCPUs (time slot of the processor) with 16Gb Ram and 100Gb Root Disk. With highly computational power, the cloud server is suitable for neural network training and testing. Stack cloud is shown in Figure 14.

\begin{figure}[!h]
	\centering
	\includegraphics[width=3.0in]{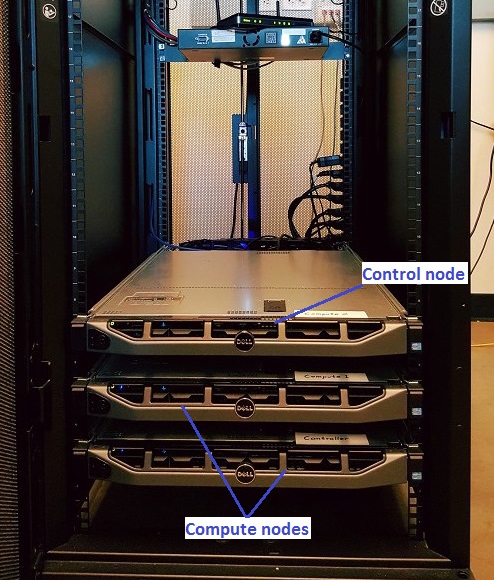}
	\caption{Cloud Node.}
	\label{webpg}
\end{figure}

\subsection{Interface collecting data from social media and owner}

We developed an android app to alert to the owner/administrator via a smart phone in Figure 15. 
\begin{figure}[!h]
	\centering
	\includegraphics[width=3.3in]{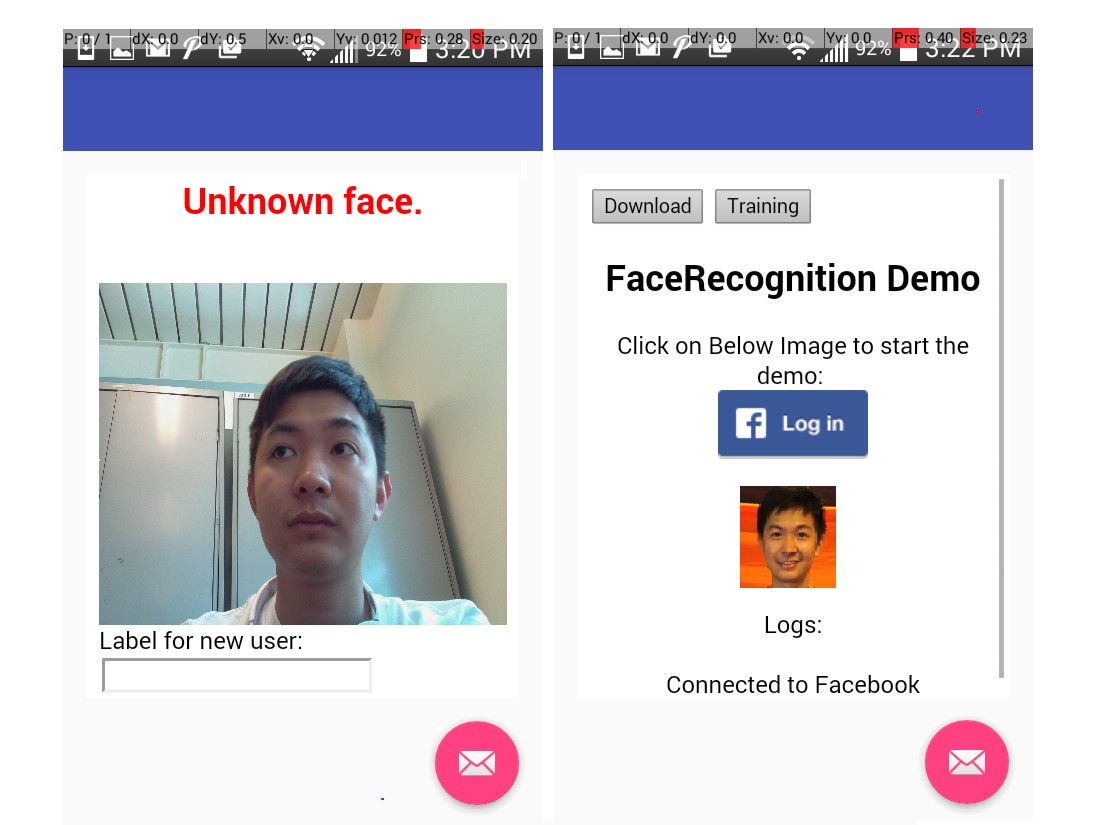}
	\caption{Android Application.}
	\label{webpg}
\end{figure}
It also is a web-based server in the cloud server, which is convenient to access anytime, anywhere from any device. The left side is interfaced with the owner, whenever someone tries to access the house, the new data will be updated in the app and on the website as well. The owner will receive a notification, which is labeled name or unknown face. Then the owner can label new users and the system will automatically retrain the classifier model with new users and give them access. On the right side, the Facebook graph API is built into the app with the necessary information from Facebook's database. We collected the tagged faces with face locations and saved to cloud storage. We also gave a permission level for different users which will protect privacy or unsupervised children in Figure 16. 
\begin{figure}[!h]
	\centering
	\includegraphics[width=3.3in]{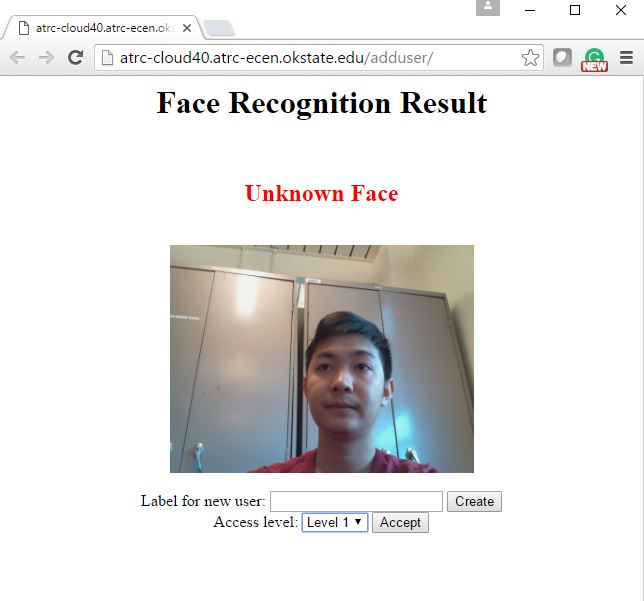}
	\caption{Web-base interface.}
	\label{webpg}
\end{figure}

For example, the guests do not have access to control the bedroom door, and the kids cannot control the dangerous electric devices or television.

\section{DISCUSSION AND FUTURE WORK}
In this paper, we introduced a new method of obtaining data for training a security system from social media and human interaction. There are several advantages of our system which can be described. 

First, we should mention that using TensorFlow is adaptive, powerful, and visualizing. Also, training time is acceptable to compare with other frameworks and faster if using distributed TensorFlow. The comparison shows in Table 1 \cite{benchmarks}.

\begin{table}[h]
	
	\centering
		%\begin{adjustbox}{max width=\textwidth}
		\resizebox{0.9\textwidth}{!}{\begin{minipage}{\textwidth}
		\begin{tabular}{|c|c|c|c|c|}
			\hline
		\textbf{Library} & \textbf{Class} & \textbf{Time(ms)} & \textbf{Forward(ms)} & \textbf{Backward(ms)}\\
			\hline
			TensorFlow & conv2d & 445 &135 & 310\\
			\hline
			CuDNN(Torch) & cudnn.Spatial & 470 &130&1148\\
			\hline
			Caffe & ConvLayer & 1935 &786&1148\\
			\hline
		
		\end{tabular}
	\end{minipage}}
	%\end{adjustbox}
	\caption{TensorFlow benchmark with GoogleNet V1 model.}
	\label{table_example}
\end{table}

There are abundant interesting projects which are leading in Artificial Intelligent and Deep Learning developed in TensorFlow with huge support from Google . In addition, computation in parallel mode will dramatically drop the training time.

By using the method mentioned in the Facenet paper \cite{schroff2015facenet}, we succeeded in reaching a robust accuracy comparison with other algorithms as in Table 2. More importantly, the accuracy is improving as long as the system is used with new data from social media and human interaction. 

\begin{table}[h]
	
	\centering
	%\begin{adjustbox}{max width=\textwidth}
	
	\begin{tabular}{|c|c|}
		\hline
		\textbf{Models} & \textbf{Accuracy} \\
		\hline
		SoF model & 0.9318 $\pm$ 0.0140	\\
		\hline	
		Facenet paper & 0.9963 $\pm$ 0.009 \\
		\hline
		Eigenface & 0.6002 $\pm$ 0.0079 \\
		\hline
		Human , cropped & 0.995 \\
		\hline

	\end{tabular}
	%\end{adjustbox}

	\caption{Face Recognition performance.}
	\label{table_example}
\end{table}

Second, collecting data from social media is also a beneficial move since social media is the largest public data such as Facebook with around 1.7 billion active users. With the publication of Facebook, we can easily collect the necessary data. We can also collect data from other social networks such as Instagram, Weibo.

One interesting direction for future work is to collect the data from the owner's smartphone such as captured images and videos and to train the network automatically. Another direction for future work is to detect fake-face by using gait speed and eye tracking.

\bibliographystyle{unsrt}
\bibliography{reference}

\end{document}